\documentclass{iau}
\usepackage{graphicx}
\usepackage{epstopdf}
\usepackage{amssymb,esint}
\usepackage{mathptmx}

\newcommand{\bea}{\begin{eqnarray}}
\newcommand{\eea}{\end{eqnarray}}
\newcommand{\beq}{\begin{equation}}
\newcommand{\eeq}{\end{equation}}

\newcommand{\apj}{\textit{ApJ}}
\newcommand{\apjl}{\textit{ApJ}}
\newcommand{\aj}{\textit{AJ}}
\newcommand{\mnras}{\textit{MNRAS}}

\newcommand{\prl}{\textit{Phys. Rev. Lett.}}

\title[Universal void density profiles] 
{Universal void density profiles from simulation and SDSS}

\author[S. Nadathur et al.]   
{S. Nadathur$^{1}$,\thanks{seshadri.nadathur@helsinki.fi} S. Hotchkiss$^{2}$, J. M. Diego$^{3}$, I. T. Iliev$^{2}$, S. Gottl\"ober$^{4}$, W. A. Watson$^{2}$ \and G.~Yepes$^5$}

\affiliation{$^1$Department of Physics, University of Helsinki and Helsinki Institute of Physics, P.O. Box 64, FIN-00014, University of Helsinki, Finland\\[\affilskip]
$^2$Department of Physics and Astronomy, University of Sussex, Falmer, Brighton, BN1 9QH, UK\\[\affilskip]
$^3$IFCA, Instituto de Fisica de Cantabria (UC-CSIC), Avda Los Castros s/n. E-39005 Santander, Spain\\[\affilskip]
$^4$Leibniz-Institute for Astrophysics, An der Sternwarte 16, D-14482 Potsdam, Germany\\[\affilskip]
$^5$ Departamento de F\'isica Te\'orica, Modulo C-XI, Facultad de Ciencias, 
Universidad Aut\'onoma de Madrid, 28049 Cantoblanco, Madrid, Spain}

\pubyear{2014}
\volume{308}  
\pagerange{1--4}
\jname{The Zeldovich Universe: genesis and growth of the cosmic web}
\editors{R. van de Weygaert, S. Shandarin, E. Saar \& J. Einasto, eds.}

\begin{document}

\maketitle

\begin{abstract}
We discuss the universality and self-similarity of void density profiles, for voids in realistic mock luminous red galaxy (LRG) catalogues from the Jubilee simulation, as well as in void catalogues constructed from the SDSS LRG and Main Galaxy samples. Voids are identified using a modified version of the {\small ZOBOV} watershed transform algorithm, with additional selection cuts.  We find that  voids in simulation are \emph{self-similar}, meaning that their average rescaled profile does not depend on the void size, or -- within the range of the simulated catalogue -- on the redshift. Comparison of the profiles obtained from simulated and real voids shows an excellent match. The profiles of real voids also show a \emph{universal} behaviour over a wide range of galaxy luminosities, number densities and redshifts. This points to a fundamental property of the voids found by the watershed algorithm, which can be exploited in future studies of voids.
\end{abstract}

\firstsection 
\section{Introduction}
Voids are recognised as particularly interesting objects for cosmology for many reasons. Of particular interest recently has been their use as probes of the expansion history via the Alcock-Paczynski test (e.g.  \cite{Lavaux:2011yh}, \cite{Hamaus:2014afa}), void-galaxy correlations (\cite{Hamaus:2013qja}, \cite{Paz:2013}) and the weak lensing signal of stacked voids (e.g. \cite{Krause:2013}, \cite{Clampitt:2014}). It has even been suggested that the integrated Sachs-Wolfe effect of voids on the CMB can be measured (\cite{Granett:2008ju}), though theoretical expectations and more recent observational results (e.g. \cite{Nadathur:2011iu}, \cite{Flender:2012wu},  \cite{Cai:2013ik}, \cite{Hotchkiss:2014}) do not support this. 

Many of these studies have assumed that voids are \emph{self-similar} objects, in particular that the density distribution in each void can be simply rescaled depending on the size of the void, and sometimes that this distribution is \emph{universal}---that is, that the rescaled void properties are independent of the properties of the tracer population in which the voids were identified or the survey redshift. The form of the density profile itself has also been the subject of study (e.g. \cite{Colberg:2005}, \cite{Ceccarelli:2013}, \cite{Nadathur:2014a}, \cite{Sutter:2013ssy}, \cite{Ricciardelli:2014}, \cite{Hamaus:2014fma}, \cite{Nadathur:2014b}), but there is a lack of consensus on the functional form of the profile as well as on the questions of self-similarity and universality. 

We make use of data from the Jubilee $N$-body simulation (\cite{Watson:2013cxa}) and SDSS DR7 galaxy catalogues to further investigate these issues. The voids are identified using a modified version of the {\small ZOBOV} watershed transform void finder (\cite{Neyrinck:2007gy}) with further selection criteria imposed to avoid spurious detections arising from Poisson noise or survey boundary contamination effects. The void catalogues from SDSS data are taken from \cite{Nadathur:2014a}, and cover voids identified in six spectroscopic volume-limited galaxy samples (\emph{dim1}, \emph{dim2}, \emph{bright1}, \emph{bright2}, \emph{lrgdim} and \emph{lrgbright}; \cite{Blanton:2004aa,Kazin:2010}), with widely varying luminosity cuts and galaxy densities. For Jubilee data we use an HOD model applied to halos on the light cone (\cite{Watson:2013cxa,Nadathur:2014b}) to obtain two mock LRG samples, referred to as JDim and JBright, whose properties match those of the \emph{lrgdim} and \emph{lrgbright}, and extract voids from those.

\section{Method}
To identify voids we use a modification of the {\small ZOBOV} algorithm (\cite{Neyrinck:2007gy}), which uses a Voronoi tessellation field estimator (VTFE) to reconstruct the galaxy density field from discrete point distribution, and then joins local minima of this field together to form voids according to the watershed algorithm. We account for the finite redshift extents of the samples and the irregular SDSS survey mask through the use of buffer particles at all survey boundaries. We restrict the merging of zones of density minima beyond linking densities $\rho_\mathrm{link}=0.3\overline\rho$ and apply a strict selection criterion on the minimum VTFE density in the void, $\rho_\mathrm{min}\leq0.3\overline\rho$. This last criterion is introduced to counter shot noise effects. Even in a uniform Poisson distribution of points, {\small ZOBOV} will always find spurious `voids' with $\rho_\mathrm{min}<\overline\rho$; however, fewer than $1\%$ of these spurious voids have $\rho_\mathrm{min}\leq0.3\overline\rho$ (\cite{Nadathur:2014a}). We have also examined stricter criteria $\rho_\mathrm{link}\leq0.2\overline\rho$ and $\rho_\mathrm{min}\leq0.2\overline\rho$, which do not materially affect our conclusions.

For each void, we define its centre to be the volume-weighted barycentre of the member galaxies of the void as identified by {\small ZOBOV}, $\mathbf{X}=\frac{1}{\sum_i V_i}\sum_i \mathbf{x}_i V_i$, where $V_i$ is the volume of the Voronoi cell of the $i$th galaxy, and the void effective radius $R_v$ to be the volume of a sphere occupying a volume equal to the sum of the Voronoi volumes of the void member galaxies.

To measure the average (stacked) galaxy density profile, we first assume self-similarity to rescale all distances from the void centre in units of the radius $R_v^i$ and construct a series of spherically symmetric radial shells each of width $\Delta$ in rescaled units. Then if mean density in the $j$th shell is estimated using the VTFE itself
 \beq
\label{eq:VTFE1}
\overline{\rho}^j = \frac{\sum_{i=1}^{N_v}\sum_{k=1}^{N_i^j} \rho_k V_k}{\sum_{i=1}^{N_v}\sum_{k=1}^{N_i^j} V_k}\,,
\eeq
where $V_k$ is the volume of the Voronoi cell of the galaxy $k$, $\rho_k$ is its density inferred from the inverse of the Voronoi volume;  the sum over $k$ runs over all galaxies in the $j$th shell of void $i$ (not only void member galaxies); and the sum over $i$ includes all voids in the stack. The error in $\overline{\rho}^j$ is estimated from jacknife samples excluding all galaxies from each of the $N_v$ voids in turn. 

\cite{Nadathur:2014b} also examined the performance of two other density estimators. The commonly used `naive' estimator, $\overline{\rho}^j = \left(\sum_{i=1}^{N_v}N_i^j/V_i^j\right)/N_v$, where $V_i^j$ is the volume of the $j$th radial shell of the $i$th void and $N_i^j$ is the number of galaxies contained within it, was shown to be systematically biased low in low density regions such as void centres. A better estimator based on Poisson statistics is $\overline{\rho}^j = \left(\sum_{i=1}^{N_v} N_i^j+1\right)/\sum_{i=1}^{N_v} V_i^j$; however, while this is optimal for simulations in a cubic box with periodic boundary conditions, it suffers from volume-leakage effects when the galaxy positions are restricted to a finite (and highly irregular) survey window, which cause it to underestimate the density at distances $r\gtrsim R_v$ for voids close to the survey edge.

\begin{figure*}[!t]
\centering
\resizebox{0.95\hsize}{!}{
\includegraphics[]{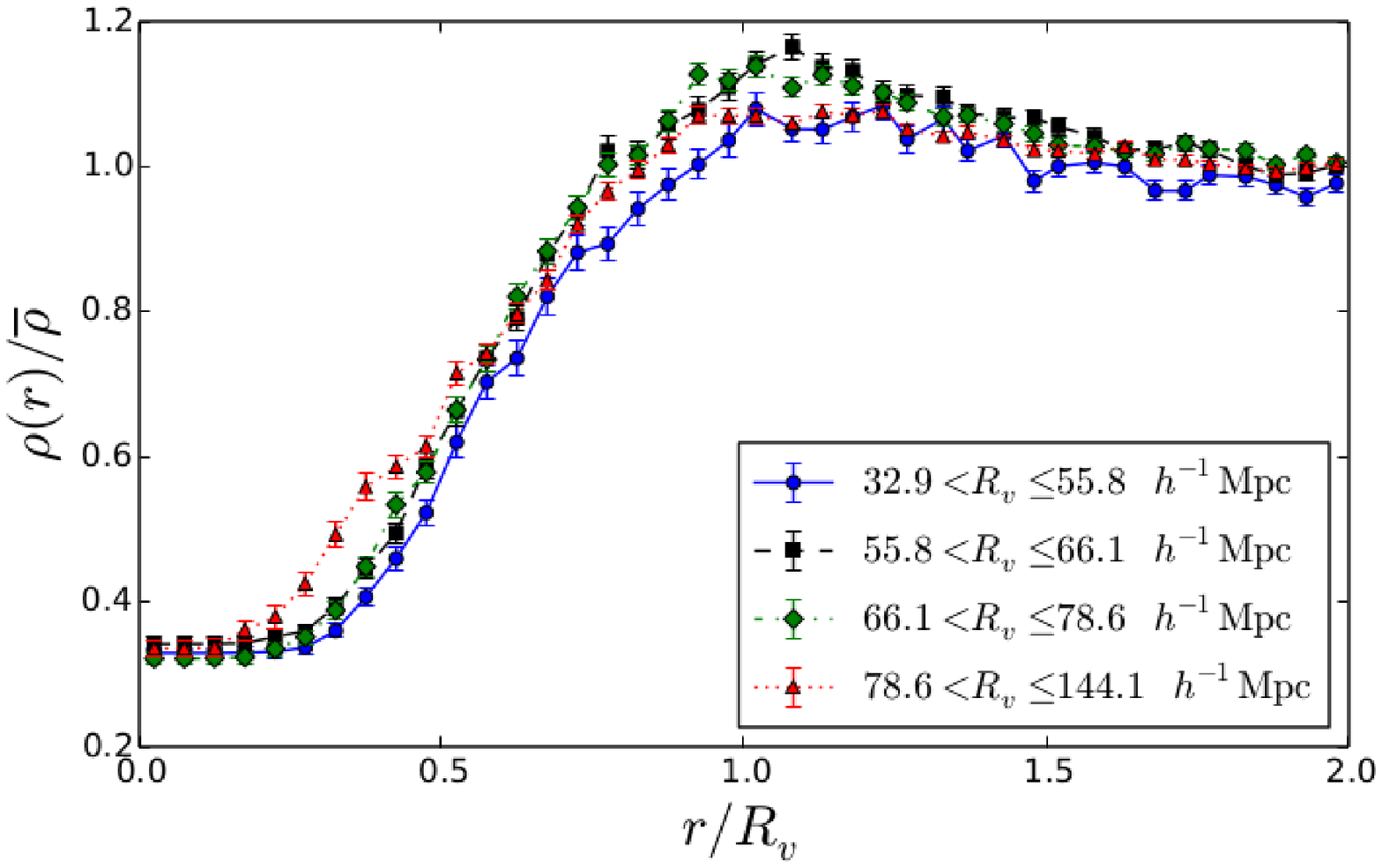}
\includegraphics[]{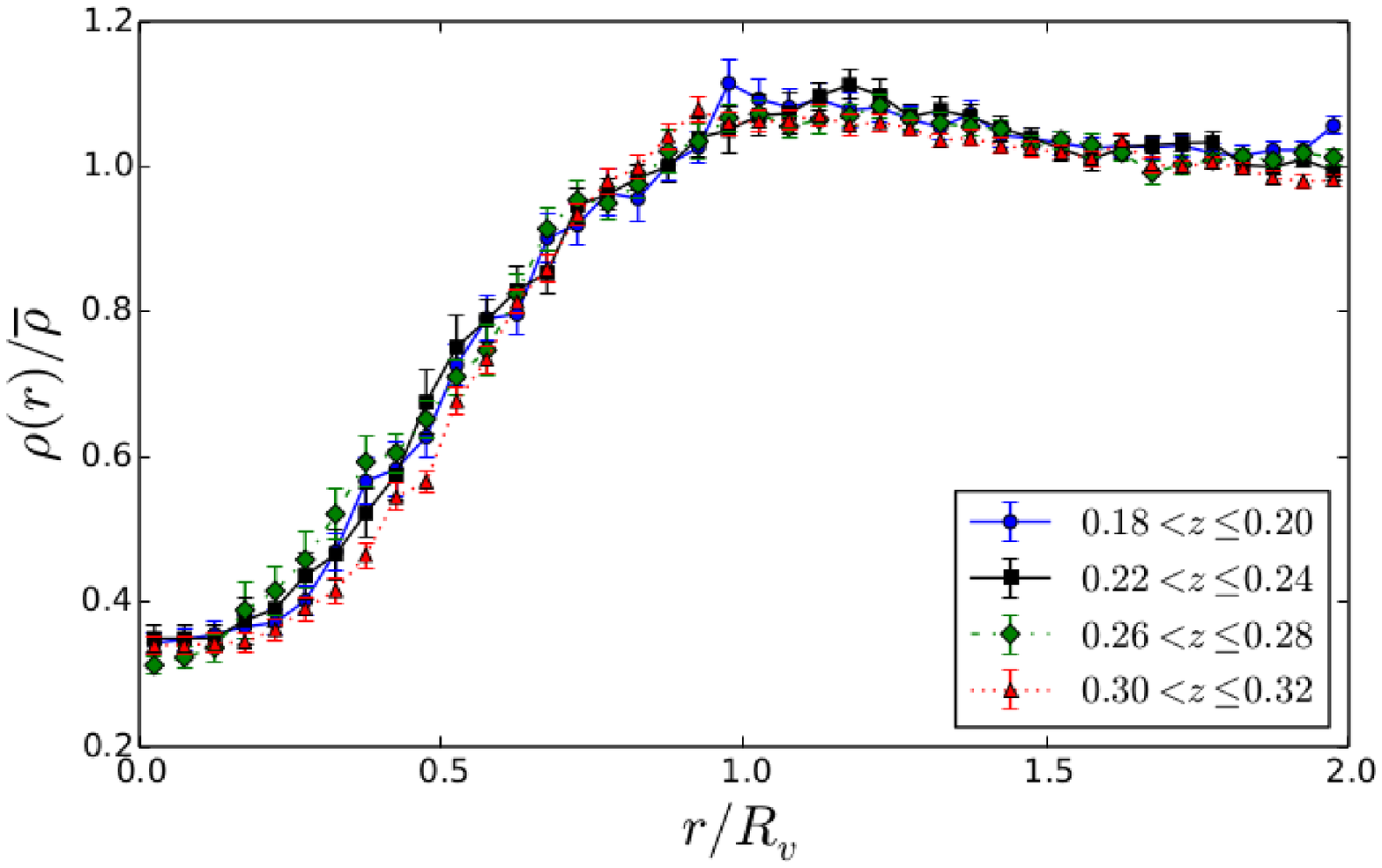}}
\caption{\emph{Left panel}: The average stacked density profiles for voids from the JDim mock catalogue, split into different quartiles of void radius $R_v$. \emph{Right}: Stacked profiles for voids in four representative redshift bins within the JDim catalogue.}
\label{fig1}
\end{figure*}

\begin{figure*}[!t]
\centering
\resizebox{0.95\hsize}{!}{
\includegraphics[]{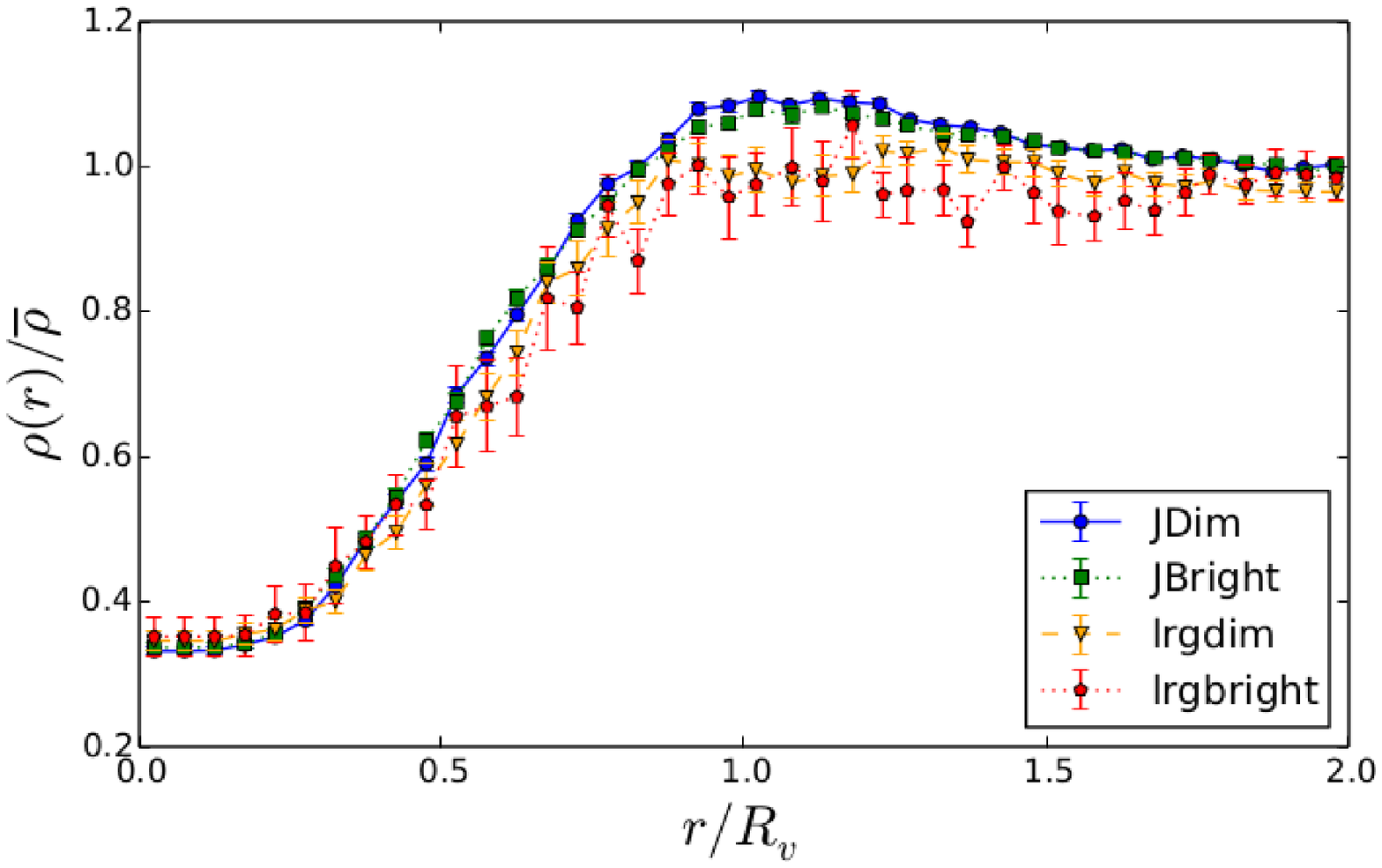}
\includegraphics[]{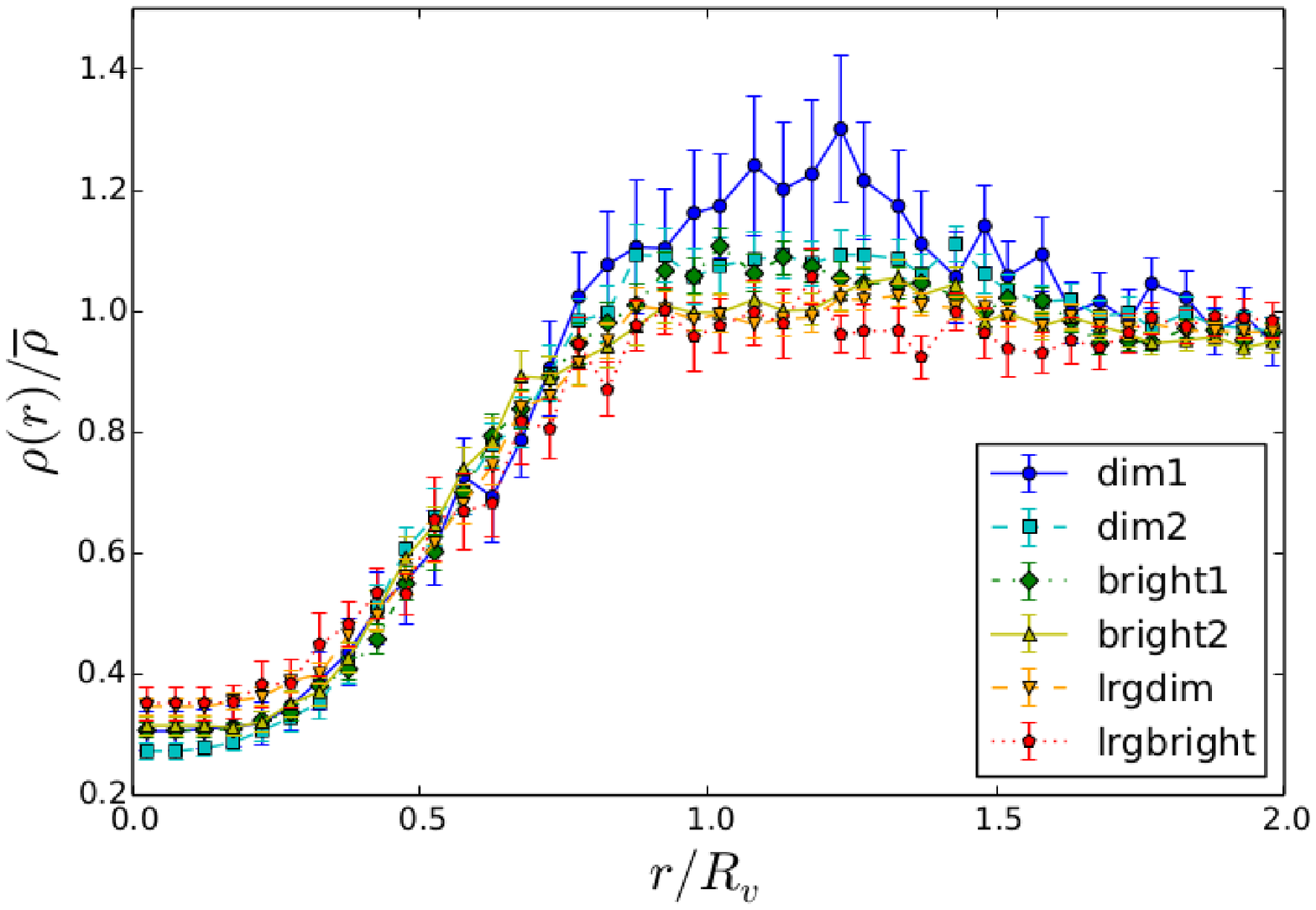}}
\caption{\emph{Left panel}: Stacked density profiles for voids from the SDSS catalogues \emph{lrgdim} and \emph{lrgbright}, compared with those from the corresponding mock catalogues JDim and JBright from Jubilee. \emph{Right}: Stacked profiles for voids from all six SDSS galaxy catalogues.}
\label{fig2}
\end{figure*}

\section{Results}
{\underline{\it Self-similarity}}. Although the stacked density estimator of eq.(\ref{eq:VTFE1}) assumes self-similarity, we can also directly test this assumption for consistency by splitting the full stack of voids into different subsets. We find that a small subset ($\lesssim5\%$) of voids in simulation consist of more than 5 different density minima amalgamated together by the watershed algorithm, and these voids shows a different density profile to the others. This is due to averaging over the internal substructure of these voids, and is also manifest in the greater distance of the barycentre from the minimum density centre. These voids can be removed from the full sample by hand, or they can be eliminated by a stricter (but somewhat arbitrary) choice of the zone merging criterion, $\rho_\mathrm{link}\leq0.2\overline\rho$ as used by \cite{Hamaus:2014fma}. When this is done, the remaining voids show no dependence of the mean stacked profile on the void radius or redshift (Fig. \ref{fig1}). We find that a simple fitting formula of the form $\frac{\rho(r)}{\overline\rho} = 1 +\delta\left(\frac{1-\left(r/r_s\right)^\alpha}{1+\left(r/r_s\right)^\beta}\right)$, with $\delta=-0.69$, $r_s = 0.81R_v$, $\alpha=1.57$ and $\beta=5.72$, provides a reasonable fit to the simulation data. This describes a one-parameter family of curves over the void size parameter $R_v$.

{\underline{\it Agreement with SDSS}}. Fig. \ref{fig2} shows that up to distances $r\sim R_v$ there is very good agreement between the density profiles from Jubilee and SDSS data. There is however a residual small difference at $r\sim R_v$, where in simulation we see an overdense compensating wall at the void edge. This may reflect some small inadequacies of the HOD modelling of LRGs in these regions, but a more likely explanation is that these are residual artefacts of the SDSS survey mask. Further detailed study of the effect of the survey mask in simulations will clarify this issue.

{\underline{\it Universality}}. In Fig. \ref{fig2}, we show the dependence of the average stacked profile on the properties of the tracer galaxies using SDSS data. These samples span a wider range of redshifts (from $z<0.05$ for \emph{dim1} to $0.16<z<0.44$ for \emph{lrgbright}), absolute magnitudes (from $M_r<-18.9$ for \emph{dim1} to $M_g<-21.8$ for \emph{lrgbright}) and mean void sizes (from $\overline{R_v}=9.6\;h^{-1}$Mpc for \emph{dim1} to $\overline{R_v}=92.8\;h^{-1}$Mpc for \emph{lrgbright}) than is available in our mock catalogues. Nevertheless, the results indicate a remarkable degree of universality in the stacked void profile across all galaxy samples. This universality is most pronounced close to the void centres; small differences are seen at the void edge. The trend seen in this edge region is consistent with the expectation that samples at lower redshift should show higher densities in the void walls simply due to greater growth of structure at late times. 


\section{Conclusions}
The density profile of voids is a subject of great interest for cosmology. Several studies have implicitly assumed the self-similarity and universality of this profile; our aim in this work has been to examine the validity of these assumptions. We find that voids in the Jubilee simulation are indeed self-similar, and that the measured profile matches that seen for voids in the SDSS DR7. In addition we have shown that void profiles from SDSS galaxy samples covering a wide range of galaxy magnitudes and number densities are also universal, being essentially indistinguishable from each other within the void interior. This significantly extends the results found from simulation. It provides a reference point for comparisons with theoretical models (\cite{Sheth:2003py}) and may prove to be a useful observational tool to constrain cosmology.

\providecommand{\noopsort}[1]{}


\begin{thebibliography}{}

\bibitem[{{Blanton}
  et~al.} {2005}]{Blanton:2004aa}
{Blanton} M.~R.  et~al., 2005, \aj, 129, 2562

\bibitem[{{Cai} et~al.} {2013}]{Cai:2013ik}
{Cai} Y.-C.,  {Neyrinck} M.~C.,  {Szapudi} I.,  {Cole} S.,    {Frenk} C.~S.,
  2013, ArXiv e-prints, 1301.6136

\bibitem[{{Ceccarelli} et~al.} {2013}]{Ceccarelli:2013}
{Ceccarelli} L.,  {Paz} D.,  {Lares} M.,  {Padilla} N.,    {Lambas} D.~G.,
  2013, \mnras, 434, 1435

\bibitem[{{Clampitt}
  et~al.} {2013}]{Clampitt:2012ub}
{Clampitt} J.,  {Cai} Y.-C.,    {Li} B.,  2013, \mnras, 431, 749

\bibitem[{{Clampitt} \&
  {Jain}} {2014}]{Clampitt:2014}
{Clampitt} J.,  {Jain} B.,  2014, ArXiv e-prints, arXiv:1404.1834

\bibitem[{Colberg et~al.} {2005}]{Colberg:2005}
Colberg J.~M.,  Sheth R.~K.,  Diaferio A.,  Gao L.,    Yoshida N.,  2005,
  \mnras, 360, 216

\bibitem[{Flender
  et~al.} {2013}]{Flender:2012wu}
Flender S.,  Hotchkiss S.,    Nadathur S.,  2013, JCAP, 1302, 013

\bibitem[{Granett
  et~al.} {2008}]{Granett:2008ju}
Granett B.~R.,  Neyrinck M.~C.,    Szapudi I.,  2008, \apj, 683, L99

\bibitem[Hamaus et al. 2014a]{Hamaus:2013qja}
Hamaus, N., Wandelt B. D., Sutter P. M.,  Lavaux G. and Warren M. S., 2014, \prl, 112, 041304

\bibitem[{{Hamaus}
  et~al.} {2014b}]{Hamaus:2014fma}
{Hamaus} N.,  {Sutter} P.~M.,    {Wandelt} B.~D.,  2014, \prl, 112, 251302

\bibitem[Hamaus et al. 2014c]{Hamaus:2014afa} 
  {Hamaus} N., Sutter P.~M., Lavaux G. and Wandelt B.~D., 2014, ArXiv e-prints,
  arXiv:1409.3580


\bibitem[{{Hotchkiss}
  et~al.} {2015}]{Hotchkiss:2014}
{Hotchkiss} S.,  {Nadathur} S.,  {Gottl{\"o}ber} S., et al.,  2015, \mnras, 446, 1321

\bibitem[{{Kazin}
  et~al.} {2010}]{Kazin:2010}
{Kazin} E.~A.  et~al., 2010, \apj, 710, 1444

\bibitem[Krause et~al. 2013]{Krause:2013}
{Krause} E.,  {Chang} T.-C.,  {Dor{\'e}} O.,    {Umetsu} K.,  2013, \apjl, 762,
  L20

\bibitem[Lavaux \& Wandelt 2012]{Lavaux:2011yh}
Lavaux G.,  Wandelt B.~D.,  2012, \apj, 754, 109

\bibitem[Nadathur \&
  Hotchkiss 2014]{Nadathur:2014a}
{Nadathur} S.,  {Hotchkiss} S.,  2014, \mnras, 440, 1248

\bibitem[{Nadathur
  et~al.} {2012}]{Nadathur:2011iu}
Nadathur S.,  Hotchkiss S.,    Sarkar S.,  2012, JCAP, 1206, 042

\bibitem[{{Nadathur}
  et~al.} {2014}]{Nadathur:2014b}
{Nadathur} S.,  {Hotchkiss} S.,  {Diego} J. M., et al.,  2014, ArXiv e-prints, arXiv:1407.1729

\bibitem[{{Neyrinck}} {2008}]{Neyrinck:2007gy}
{Neyrinck} M.~C.,  2008, \mnras, 386, 2101

\bibitem[{{Paz} et~al.} {2013}]{Paz:2013}
{Paz} D.,  {Lares} M.,  {Ceccarelli} L.,  {Padilla} N.,    {Lambas} D.~G.,
  2013, \mnras, 436, 3480

\bibitem[{{Ricciardelli} et~al.} {2014}]{Ricciardelli:2014}
{Ricciardelli} E.,  {Quilis} V.,    {Varela} J.,  2014, \mnras, 440, 601

\bibitem[{Sheth \& van~de
  Weygaert} {2004}]{Sheth:2003py}
Sheth R.~K.,  van~de Weygaert R.,  2004, \mnras, 350, 517

\bibitem[{Sutter et~al.} {2014}]{Sutter:2013ssy}
{Sutter} P.~M.,  {Lavaux} G.,  {Hamaus} N.,  {Wandelt} B.~D.,  {Weinberg}
  D.~H.,    {Warren} M.~S.,  2014, \mnras, 442, 462

\bibitem[Watson et~al., 2014]{Watson:2013cxa}
{Watson} W.~A.  et~al., 2014, \mnras, 438, 412



\end{thebibliography}
\end{document}